# Self-organising Roles in Agile Globally Distributed Teams

Sherlock A. Licorish
*SERL, School of Computing and Mathematical Sciences*
*Auckland University of Technology*
*Auckland, New Zealand*
*slicoris@aut.ac.nz*

Stephen G. MacDonell
*Department of Information Science*
*University of Otago, Dunedin*
*New Zealand*
*stephen.macdonell@otago.ac.nz*

## Abstract

*The ability to self-organise is posited to be a fundamental requirement for successful agile teams. In particular, self-organising teams are said to be crucial in agile globally distributed software development (AGSD) settings, where distance exacerbates team issues. We used contextual analysis to study the specific interaction behaviours and enacted roles of practitioners working in multiple AGSD teams. Our results show that the teams studied were extremely task focussed, and those who occupied team lead or programmer roles were central to their teams' self-organisation. These findings have implications for AGSD teams, and particularly for instances when programmers – or those occupying similar non-leadership positions – may not be willing to accept such responsibilities. We discuss the implications of our findings for information system development (ISD) practice.*

## 1. INTRODUCTION

Contemporary thinking regarding ISD projects recognises that a range of human and social factors, rather than technical issues, have the most substantial influence on project outcomes. Among these factors, the matching of practitioners to certain roles has been shown to benefit task performance in projects (Acuna et al. 2006b). Such results imply that particular ISD activities require specific work behaviours, and individuals who demonstrate appropriate levels of those behaviours would perform most effectively in the corresponding roles. The consequence of such findings is that role assignment should be conducted, and actively managed, in relation to individuals' specific capabilities, characteristics and behaviours.

Agile software development methods such as Extreme Programming (XP), Adaptive Software Development (ASD), and SCRUM challenge this thinking, as these methods emphasise the need for self-organisation and flexible team role assignment (Pressman 2009). Empirical studies of self-organising agile teams have found evidence that project team members do indeed adopt various roles, as needed, to facilitate self-organisation. Hoda et al. (2010), for instance, identified the roles of mentor, translator, champion, coordinator, promoter and terminator, and saw them assumed at various times by different team members so that progress in projects could be sustained. That said, while such flexible adoption of roles is likely to be evident and considered necessary in agile ISD contexts (Hoda et al. 2010), other prior work has noted that it is rarely achieved (Moe et al. 2008). Moreover, there have been few investigations of issues of expertise, role assignment, role adoption and self-organisation in globally distributed development contexts beyond those related to open source software (OSS) (Crowston et al. 2007). This is despite the relevance of such phenomena in AGSD settings, with their inherently limited opportunities for timely communication among dispersed team members (Serce et al. 2009).

While those studying OSS teams have provided insights into the way such teams self-organise, these teams utilise different processes to those used in commercial settings. In most OSS contexts individuals contribute voluntarily to projects for reasons often associated with personal interest and ideological commitment (Oreg and Nov 2008). In contrast, developers' motivations in commercial projects are likely to be divergent given likely immediate rewards (e.g., financial remuneration). We have therefore extended our previous work that utilised psycholinguistics (Licorish and MacDonell 2013), and have employed content analysis techniques to examine the specific interaction behaviours and enacted roles adopted by those occupying a range of formally assigned roles while working in multiple commercial AGSD teams.

Through this extension study we provide explanations for the way agile teams actually self-organise, along with recommendations for agile team composition and project governance. In the next section we survey related work and outline our research questions. Our research setting is then outlined, prior to the presentation of our results. Discussion of our main findings follows. Finally, we conclude this work and consider our study's limitations.

## 2. BACKGROUND AND RELATED WORK

### Agile Globally Distributed Software Development (AGSD)

Geographically distributed work is becoming ubiquitous due to globalisation, and this trend has found favour in numerous ISD organisations (Bird et al. 2009). For instance, India's software industry grew between 30% and 40% annually for the ten year period ending in 2004 due to their involvement in global software ventures (Arora and Gambardella 2005). Driven by the availability of cheaper hardware, affordable software development talent pools, increased access to communication infrastructure and technologies and the need to reduce time-to-market, many software companies have expanded their operations to both leverage and reach global contexts. In keeping with this expansion, these companies are employing AGSD approaches (Layman et al. 2006).

Although many success stories have been reported regarding the implementation of agile methodologies in AGSD contexts (Layman et al. 2006), this approach has also been reported to be quite challenging (Kamaruddin et al. 2012). In particular, team member dispersion in AGSD has been shown to reduce the opportunities for informal (and face-to-face) communication (Cataldo et al. 2007). This dispersion has also been shown to detrimentally affect project oversight and monitoring, and temporal distance has been reported to have a negative impact on team culture and trust (Lee and Yong 2010).

Due to the way AGSD teams operate in a distributed manner, individual team members must often rely heavily on communication technologies to support their team processes (Bachmann and Bernstein 2009). Of the potential risks that arise in AGSD, project communication in particular is often critical to teams' performance (Herbsleb and Mockus 2003). Given that team communication is often recorded for persistence in AGSD settings, such communications form a source that could provide novel details of the ISD process (Abreu and Premraj 2009). The rationale for project decisions, pointers for how AGSD teams work, insights into the way such teams collaborate, and rich information on AGSD team dynamics are stored in distributed software teams' communication logs. Thus, these logs could provide invaluable knowledge-bases relating to AGSD, as has been demonstrated (Bachmann and Bernstein 2009; Abreu and Premraj 2009; Herbsleb and Mockus 2003).

The work reported here uses a sample of such artefacts to examine the specific interaction behaviours and enacted roles adopted by those occupying a range of formally appointed roles while they were working in multiple AGSD teams. We introduce the principles and theories around team roles and self-organising teams to provide the theoretical basis for our inquiry in the following subsection.

### Principles of Team Roles and Self-organising Teams

According to theories in the psychology and management disciplines, social and team role principles may be used to characterise individual behaviours and their personal interactions in teams, and each individual's behavioural style is correlated with their occupation of specific team roles (Belbin 2002). Meredith Belbin proposed a model for assigning participants to roles during team work, after nearly a decade of observation embedded in personality psychology in five different countries (Belbin 2002). During his observations, Belbin observed that individuals in teams occupied nine distinct roles: Implementer (IM), Co-ordinator (CO), Shaper (SH), Plant (PL), Resource Investigator (RI), Monitor Evaluator (ME), Team Worker (TW), Completer/Finisher (CF) and Specialist (SP). Belbin asserts that in successful teams, these roles are performed by various team members. Of course, individuals may also enact very divergent roles to those that are nominally assigned at project initiation (Licorish and MacDonell 2013).

In relation to ISD groups or departments, roles may relate to the specific software process or methodology being utilised. For instance, a group that has adopted XP will likely define roles such as programmer, tester, coach and so on (Highsmith 2004). In addition, roles may sometimes be performed arbitrarily by team members in which case these members must possess a level of general competency in many roles (Gorla and Lam 2004). Thus in this context, role arrangement and competency requirements for individual ISD-related roles may be subject to specific and dynamic organisational requirements.

Given the emphasis on self-organising teams that is evident in some Agile approaches (Pressman 2009), the question "how do teams self-organise?" has been the focus of both software engineering (SE) and IS research. In fact, the ability to self-organise has been purported to be one of the key determinants of agile teams' success (Hoda et al. 2010). In order to self-organise, various team members are said to adopt informal roles beyond (or perhaps instead of) their assigned roles as the need arises (Pressman 2009). However, Moe et al. (2008) noted that the process of self-organisation is actually quite complex, and so may not suit all ISD contexts. Their ethnographic study in Norway of novice agile practitioners revealed that team members displayed minimal internal autonomy and were rarely willing to assume roles other than those that matched their specialised competencies. These findings may be contrasted with those of Hoda et al. (2010), who found that agile developers in India and New Zealand operated more fluidly across assigned and non-assigned roles. This signals a need for further research, and particularly, explorations that may provide insights for confirmation of these alternative views. The questions outlined in the following subsection are aimed at addressing this need.

### Research Questions

The divergence in findings evident in the studies outlined above suggests a need for additional research, to provide further understanding of how different roles are actually enacted by those assigned to specific roles during AGSD

teamwork. We address this research opportunity by answering the following research questions:

RQ1. What interaction behaviours are exhibited by self-organising teams?

RQ2. How are roles enacted during agile globally distributed software development?

## 3. RESEARCH SETTING

To address the research questions just specified we conducted a single field study in which we examined artefacts and messages extracted from a specific release (1.0.1) of Jazz based on the IBM$^R$ Rational$^R$ Team Concert$^{TM}$ (RTC)[1]. Jazz, created by IBM, is a fully functional environment for developing software systems and managing the entire software development process (Frost 2007). The software includes features for work planning and traceability, software builds, code analysis, bug tracking and version control, all captured in one system. Changes to source code in the Jazz environment are only allowed as a consequence of work items (WIs) being created beforehand, in the form of a bug report, a new feature request or a request to enhance an existing feature. The Jazz repository comprised a large amount of process data from development and management activities carried out across the USA, Canada and Europe. Jazz teams use the "Eclipse-way" approach for guiding the software development process. This approach outlines iteration cycles that are six weeks in duration, comprising planning, development and stabilizing phases, where practitioners share features and requirements are constantly evolving – practices aligned with agile methods' thinking (Pressman 2009), even though the iterations themselves may be longer than is typical. Builds are executed after project iterations. All information for the software process is stored in a server repository that is accessible through a web-based or Eclipse-based RTC client interface.

**Data and Sample Selection**

We created a Java program to leverage the Jazz Client API to extract information along with development and communication artefacts from ten teams (shown in Table 1) from the Jazz repository. This included: Work Items (WIs) and history logs, representing project management and development tasks; Project Workspaces, representing multiple team areas and including information on team memberships and roles; and Messages, representing practitioner dialogues and communication around project WIs.

The selected project artefacts related to 1201 development tasks, involving 394 contributors belonging to five different roles (described below), and 5563 messages exchanged around the 1201 tasks. As the data were analysed, it became clear that the ten cases selected were representative of those in the repository, as we reached saturation (Glaser and Strauss 1967) after analysing the third project case. Additionally, we used social network analysis (SNA) to explore the teams' communications and noted that all ten teams had similar profiles for network density (between 0.02 and 0.14) and closeness (between 0 and 0.06). Formal statistical testing for significant differences in *in-degree* measures also confirmed that the teams were relatively homogenous, $X^2 = 13.182$, $p = 0.155$ (Kruskal-Wallis test result).

In earlier work we used psycholinguistics to study the way these IBM Rational Jazz practitioners enacted various roles, expressed attitudes and shared competencies to successfully self-organise in their global project (Licorish and MacDonell 2013). Among our findings, we uncovered that practitioners enacted a range of roles depending on their teams' task portfolio; and that team leaders were most critical to self-organisation. The psycholinguistic approach was applied in a top-down fashion, where the categories of language codes were pre-determined and granular, considering the use of isolated words. We anticipated that a more exploratory, bottom-up approach focused on phrases might provide different insights into the way AGSD teams self-organise. We therefore studied all of the messages exchanged by three of the ten teams using a contextual analysis approach, to examine the interaction behaviours and enacted roles adopted by those that were formally assigned a range of roles during AGSD (see teams P1, P7 and P8 highlighted in Table 1). These three teams were deliberately selected as they were charged with addressing different forms of software tasks, and so, we anticipated to also reveal variations in the way teams work given their portfolio of features.

The role information extracted from the repository is as follows: Team leads (or component leads) are responsible for planning and executing the architectural integration of components; Admins are responsible for the configuration and integration of artefacts; Project managers (PMC) are responsible for project governance; those occupying the Programmer (contributor) role contribute code to features; and finally, those who occupied more than one of these roles were labelled Multiple. We used these practitioners' roles as our unit of analysis, we made comparisons of interaction behaviours across roles in individual teams, and we also conducted assessments across various task types.

**Analysis**

We studied the messages contributed by practitioners in P1, P7 and P8 using a directed content analysis (CA) approach, employing a hybrid classification scheme adapted from related prior work. The classifications schemes of Henri and Kaye (1992) and Zhu (1996) are particularly applicable to the work undertaken in this research because of their treatment of teams' interaction – the study of which should reveal the reason for team members' communication and expose their behaviours and enacted roles (further synthesised through the Belbin (2002) model). Additionally, these instruments were repeatedly validated by others. Use of a directed CA approach is appropriate when there is scope to extend or complement existing theories around a phenomenon (Hsieh and Shannon 2005), and so suited our further explorations of practitioners' roles. The directed content analyst approaches the data analysis process using existing theories to identify key concepts and definitions as initial coding categories. In our case, we used theories examining textual interaction (Henri and Kaye 1992; Zhu 1996) to inform our initial categories (Scales 1-8 in Table 2). Should existing theories prove insufficient to capture all relevant insights during preliminary coding, new categories and subcategories should be created (Hsieh and Shannon 2005). Both authors of this work and two other trained coders categorized 5% of the communications (randomly chosen) in a preliminary coding phase, and found that some aspects of Jazz practitioners' interaction behaviours were not captured by the first version of our protocol (e.g., Instructions and Gratitude – refer to Table 2). Coders were provided with guidelines for administering, scoring, and interpreting the coding scheme, including examples of messages that were coded under each category. During the pilot coding exercise we also found that Jazz practitioners communicated multiple ideas in their messages. Thus, we segmented the communication using the sentence as the unit of analysis. We extended the protocol by deriving new scales directly from the pilot Jazz data (see Scales 9 to 13 in Table 2), after which the first author and the two trained coders recoded all the messages. Duplicate codes were assigned to utterances that demonstrated multiple forms of interaction, and all coding differences were discussed and resolved by consensus. We achieved an 81% inter-rater agreement among the three coders using Holsti's (Holsti 1969) coefficient of reliability measurement (C.R). This represents excellent agreement among the coders.

**Table 1**. Summary Statistics for the Selected Jazz Project Teams

| Team ID | Task (WI) Count | Software Tasks | Total Contributors – Roles | Total Messages | Period (days) – Iterations |
|---|---|---|---|---|---|
| P1 | 54 | User Experience – tasks related to UI development | 33 – 18 programmers, 11 team leads, 2 project managers, 1 admin, 1 multiple roles | 460 | 304 - 04 |
| P2 | 112 | User Experience – tasks related to UI development | 47 – 24 programmers, 14 team leads, 2 project managers, 1 admin, 6 multiple roles | 975 | 630 - 11 |
| P3 | 30 | Documentation – tasks related to Web portal documentation | 29 – 12 programmers, 10 team leads, 4 project managers, 1 admin, 2 multiple roles | 158 | 59 - 02 |
| P4 | 214 | Code (Functionality) – tasks related to development of application middleware | 39 – 20 programmers, 11 team leads, 2 project managers, 2 admins, 4 multiple roles | 883 | 539 - 06 |
| P5 | 122 | Code (Functionality) – tasks related to development of application middleware | 48 – 23 programmers, 14 team leads, 4 project managers, 1 admin, 6 multiple roles | 539 | 1014 - 17 |
| P6 | 111 | Code (Functionality) – tasks related to development of application middleware | 25 – 11 programmers, 9 team leads, 2 project managers, 3 multiple roles | 553 | 224 - 13 |
| P7 | 91 | Code (Functionality) – tasks related to development of application middleware | 16 – 6 programmers, 7 team leads, 1 project manager, 1 admin, 1 multiple roles | 489 | 360 - 11 |
| P8 | 210 | Project Management – tasks under the project managers' control | 90 – 29 programmers, 24 team leads, 6 project managers, 2 admins, 29 multiple roles | 612 | 660 - 16 |
| P9 | 50 | Code (Functionality) – tasks related to development of application middleware | 19 – 10 programmers, 3 team leads, 4 project managers, 2 multiple roles | 254 | 390 - 10 |
| P10 | 207 | Code (Functionality) – tasks related to development of application middleware | 48 – 22 programmers, 12 team leads, 2 project managers, 1 admin, 11 multiple roles | 640 | 520 - 11 |
| ∑ | **1201** | - | **394 contributors,** comprising 175 programmers, 115 team leads, 29 project managers, 10 admins, 65 multiple roles | **5563** | - |

**Table 2**. Coding Categories for Measuring Interaction

| Scale | Category | Characteristics and Example |
|---|---|---|
| 1 | Type I Question | Ask for information or requesting an answer – "Where should I start looking for the bug?" |
| 2 | Type II Question | Inquire, start a dialogue - "Shall we integrate the new feature into the current iteration, given the approaching deadline?" |
| 3 | Answer | Provide answer for information seeking questions - "The bug was noticed after integrating code change 305, you should start debugging here." |
| 4 | Information sharing | Share information – "Just for your information, we successfully integrated change 305 last evening." |
| 5 | Discussion | Elaborate, exchange, and express ideas or thoughts – "What is most intriguing in re-integrating this feature is how refactoring reveals issues even when no functional changes are made." |
| 6 | Comment | Judgemental – "I disagree that refactoring may be considered the ultimate test of code quality." |
| 7 | Reflection | Evaluation, self-appraisal of experience – "I found solving the problems in change 305 to be exhausting, but I learnt a few techniques that should be useful in the future." |
| 8 | Scaffolding | Provide guidance and suggestions to others – "Let's document the procedures that were involved in solving this problem 305, it may be quite useful." |
| 9 | Instruction/ Command | Directive – "Solve task 234 in this iteration, there is quite a bit planned for the next." |
| 10 | Gratitude/ Praise | Thankful or offering commendation – "Thanks for your suggestions, your advice actually worked." |
| 11 | Off task | Communication not related to solving the task under consideration – "How was your weekend?" |
| 12 | Apology | Expressing regret or remorse – "Sorry for the oversight and the failure this has caused." |
| 13 | Not Coded | Communication that does not fit codes 1 to 12. |

# 4. RESULTS

**Artefacts and Codes**

The artefacts selected for the three teams together comprised 355 tasks and 1561 messages, with 139 contributors working across the three teams (comprising 107 distinct members) (refer to Table 1). All of the 1561 messages were coded using the directed CA approach outlined above. From the total 1561 messages that were coded, 5218 utterances were recorded for the three teams (P1 = 1165 codes, P7 = 1770 codes and P8 = 2283 codes). We provide other descriptive statistics for the three teams in Table 3.

**Table 3**. Mean Team Measures for Messages, Tasks, Contributors and Codes

| Team ID | Messages/ Task | Tasks/ Contributor | Messages/ Contributor | Codes/ Message |
|---|---|---|---|---|
| P1 (UE) | 8.5 | 1.6 | 13.9 | 2.5 |
| P7 (Code) | 5.4 | 5.7 | 30.6 | 3.6 |
| P8 (PM) | 2.9 | 2.3 | 6.8 | 3.7 |

**Interaction Behaviours**

Figure 1 shows the distribution of the aggregated interaction behaviours (from the 5218 derived codes) that were exhibited for the three teams P1, P7 and P8. It is evident that Information sharing (2452 codes), Discussion (598 codes), Scaffolding (590 codes) and Comments (383 codes) were the most dominant behaviours during Jazz practitioners' discourses. Additionally, Apology type communication (17 codes) was rarely observed, and only a few utterances were not matched to a category (Not Coded = 7 codes). Figure 1 also shows that Type I Questions (104 codes), Gratitude (97 codes) and Off task utterances (107 codes) had comparatively low usage and were relatively even in number. A similar pattern is evident for Type II Questions (255 codes), Answers (257 codes) and Instructions (200 codes). The number of codes for Reflection (151 codes) was slightly lower again.

These codes were then separated and grouped according to the three project teams, leading to the results presented in Figure 2. Given the relatively low numbers of codes assigned to the Apology and Not Coded categories these were omitted from the figure. Figure 2 (a) shows that although there were differences in the number of codes (as for number of messages) contributed by the three teams, the pattern of results for all of the three teams remained similar to those in Figure 1. We normalised these codes across the three teams by using percentages in Figure 2 (b); here the results also confirmed that the pattern in Figure 1 is maintained.

A Pearson Chi-square test was conducted to ascertain whether the differences observed in the visualisations shown in Figure 2 were statistically significant. This statistical procedure is appropriate when the distributions comprise frequency data, as is the case for the codes that were obtained for P1, P7 and P8 through the directed CA process (Sharp 1979). Additionally, given that the data analysed is categorical, the Chi-square test is the statistical procedure of choice. Further, with the exception of the Not Coded category (as only seven codes were recorded for this category), all the categories comprised a sample size that was substantially more than ten (as is assumed if utilising a Chi-square test) (Sharp 1979). This Chi-square result was statistically significant, $X^2 (24) = 255.523$, $p < 0.001$. However, the effect size for this finding, assessed using Cramer's V, was small, 0.221 (Cohen 1988). This result suggests that, overall, practitioners did not contribute consistently across interaction categories for the three teams, though, this difference is of limited practical significance (Kampenes et al. 2007).

**Interaction Behaviours for Individual Roles**

The interaction behaviours for the different roles were then grouped, and their (highly skewed) distributions are shown in Figure 3. These results indicate that team leads and programmers were dominant in their teams, across all interaction behaviour categories. Given that these member types also had the largest membership in Table 1 we further examine their performance in the three teams in Table 4. The pattern of results in Table 4 confirms the team leads' and programmers' dominance in their individual teams. We normalise these codes in Table 5 in order to examine practitioners' contributions in their given role in relation to their team's performance. Table 1 shows that for P1 (UE) there were 33 members in total; 29 members belonged to the programmer or team lead roles and 4 members occupied the other roles. Tables 4 and 5 show that, apart from the measures for Answers and Comments for the project managers on the UE team, all other measures for those that occupied multiple, admin and project manager roles were below the team's average. This suggests that the team leads and programmers indeed dominated this team. The corresponding results for P7 (Code) were substantially lower, with those occupying multiple, admin and project manager roles contributing much lower than their mean team contribution.

A slightly different pattern is observed for P8 (PM). Table 1 shows that P8 had 90 members; 53 members were assigned to the programmer and team lead roles and 37 members were assigned to the other roles. Tables 4 and 5 show that Questions were asked evenly across the five roles for the PM team; however, those occupying multiple roles provided the fewest Answers and Information (when compared to the mean project measures). Discussion, Comment, Reflection and Scaffolding were contributed by those occupying the programmer, team lead and project manager roles. Instructions were also provided by those occupying these roles, with those in the admin role also contributing this form of utterance. In contrast, those occupying multiple roles expressed the most Gratitude, and programmers communicated most Off task. We explore these findings in relation to theory next.

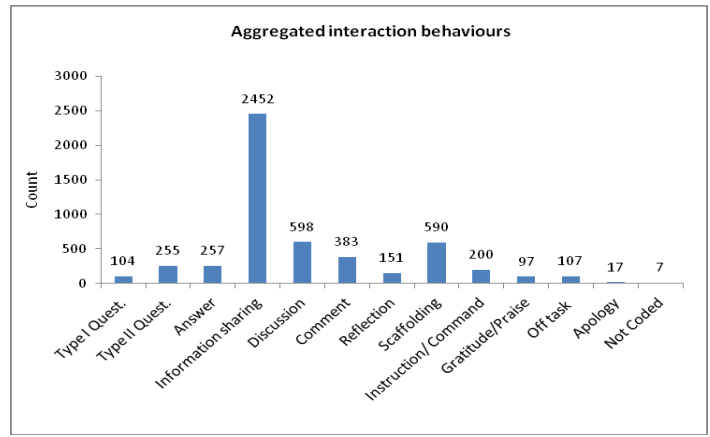

Figure 1: Aggregated interaction behaviours

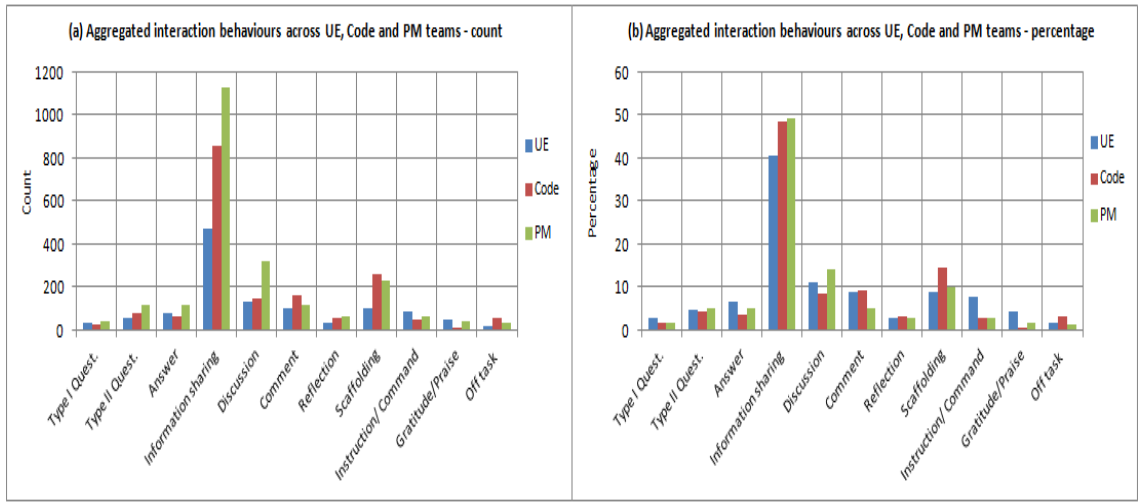

Figure 2: Aggregated interaction behaviours for the three project teams

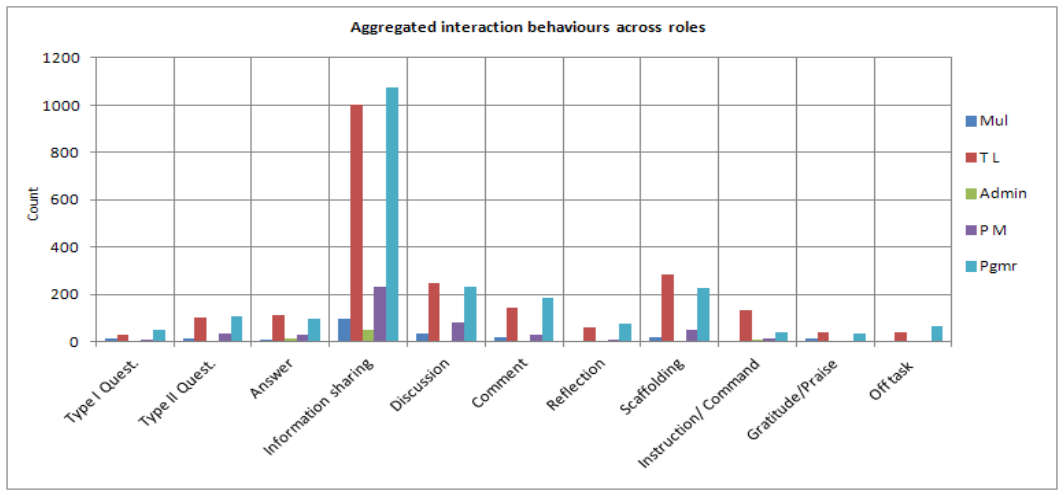

Figure 3: Aggregated interaction behaviours across roles

Table 4. Counts of Interaction Behaviours for Individual Roles in the UE, Code and PM Teams

| Category | UE | | | | | Code | | | | | PM | | | | |
|---|---|---|---|---|---|---|---|---|---|---|---|---|---|---|---|
| | Mul | T L | Admin | P M | Pgmr | Mul | T L | Admin | P M | Pgmr | Mul | T L | Admin | P M | Pgmr |
| Type I Quest. | 0 | 13 | 0 | 0 | 21 | 0 | 10 | 0 | 1 | 17 | 11 | 8 | 2 | 8 | 13 |
| Type II Quest. | 0 | 31 | 0 | 1 | 25 | 0 | 30 | 0 | 1 | 47 | 11 | 40 | 2 | 32 | 35 |
| Answer | 0 | 33 | 3 | 6 | 34 | 0 | 29 | 2 | 0 | 34 | 9 | 51 | 7 | 22 | 27 |
| Info sharing | 1 | 202 | 1 | 6 | 261 | 1 | 299 | 4 | 0 | 551 | 94 | 499 | 44 | 226 | 263 |
| Discussion | 0 | 71 | 0 | 5 | 53 | 0 | 54 | 0 | 1 | 92 | 36 | 123 | 4 | 73 | 86 |
| Comment | 0 | 50 | 0 | 7 | 45 | 0 | 50 | 0 | 2 | 113 | 21 | 45 | 5 | 20 | 25 |
| Reflection | 0 | 16 | 0 | 0 | 15 | 0 | 14 | 0 | 0 | 40 | 4 | 31 | 1 | 10 | 20 |
| Scaffolding | 0 | 62 | 0 | 1 | 40 | 0 | 122 | 1 | 0 | 137 | 20 | 101 | 4 | 50 | 52 |
| Instruction/Command | 0 | 71 | 0 | 0 | 17 | 0 | 32 | 0 | 0 | 17 | 3 | 31 | 6 | 15 | 8 |
| Gratitude/Praise | 0 | 26 | 0 | 0 | 23 | 0 | 4 | 0 | 0 | 4 | 16 | 11 | 0 | 2 | 9 |
| Off task | 0 | 13 | 0 | 0 | 7 | 0 | 20 | 0 | 0 | 36 | 3 | 5 | 0 | 0 | 23 |

KEYS:- Mul = Multiple, TL = Team lead, PM = Project manager, Pgmr = Programmer

Table 5. Counts of Interaction Behaviours and Team Average for the UE, Code and PM Teams

| Category | UE | | Code | | PM | |
|---|---|---|---|---|---|---|
| | Code Count | Team Average | Code Count | Team Average | Code Count | Team Average |
| Type I Quest. | 34 | 1.0 | 28 | 1.8 | 42 | 0.5 |
| Type II Quest. | 57 | 1.7 | 78 | 4.9 | 120 | 1.3 |
| Answer | 76 | 2.3 | 65 | 4.1 | 116 | 1.3 |
| Information sharing | 471 | 14.3 | 855 | 53.4 | 1126 | 12.5 |
| Discussion | 129 | 3.9 | 147 | 9.2 | 322 | 3.6 |
| Comment | 102 | 3.1 | 165 | 10.3 | 116 | 1.3 |
| Reflection | 31 | 0.9 | 54 | 3.4 | 66 | 0.7 |
| Scaffolding | 103 | 3.1 | 260 | 16.3 | 227 | 2.5 |
| Instruction/Command | 88 | 2.7 | 49 | 3.1 | 63 | 0.7 |
| Gratitude/Praise | 51 | 1.6 | 8 | 0.5 | 38 | 0.4 |
| Off task | 20 | 0.6 | 35 | 3.5 | 21 | 0.3 |

## 5. DISCUSSION AND IMPLICATIONS

**Self-organising Teams' Interaction Behaviours**

Self-organising teams engage in large amounts of Information sharing, Discussion and Scaffolding. Results in this work shows that this pattern of behaviour was maintained for all three teams regardless of their task portfolio and the distribution of team roles. Notwithstanding the effect of possible sampling on these results, these findings suggest that membership of AGSD teams may need to possess significant diversity of skills in order to succeed. For instance, although there were few Questions aimed at seeking help or guidance among teams overall, there was great willingness among team members to provide direction to those team members that were less aware. The need for highly skilled and willingly communicative practitioners may have implications for instances when such individuals are not available.

Our findings show that self-organising teams did not ask many Questions, and rarely communicated Off task. From a role perspective, the results suggest that although there was some evidence for social behaviours, task centred behaviours dominated the Jazz project environment. In fact, there was also some evidence in favour of debate centred roles. Thus, all roles may indeed be necessary during AGSD, as was previously posited for group work in general (Belbin 2002). Although task-focused individuals are most productive, evidence of debate related activities may not threaten project success. These behaviours may complement each other. In fact, the level of debate observed may also support the view that Jazz members were willing to critique one another's work, and were keenly involved in their teams' processes. Such a balance may be necessary to enhance innovativeness and critical evaluation among group members (Tjosvold 2008).

Findings in this work also suggest that significant levels of mentoring occur during ISD projects (evident in the results for Scaffolding), and teams with members who are willing to provide coverage for others (through starting dialogue about project features, answering questions and sharing information) as against working alone may be very useful to team performance. In AGSD, it does appear that gains can be made if team members embrace a "teamwork" mindset. Also, our findings for Scaffolding (i.e., recommendation for documenting procedures for later use by others) suggest that some documentation may be useful for maintaining team knowledge (Boehm and Turner 2003). This activity may also reduce the burden and distraction of new team members on their more experienced and productive colleagues.

**Assigned and Enacted Roles**

Findings in this work show that team members adopted various roles in order to self-organise, but that those who were formally assigned to the team lead and programmer roles were most important to their teams' self-organising processes. These two groups of practitioners were integrally involved with team organisation and task assignment (e.g., see measures for Answers and Instruction), responsibilities typical of those occupying Belbin's (2002) CO and SH roles. These findings are in contrast to those that have been uncovered for OSS teams, where self-assignment was most evident (Crowston et al. 2007), confirming that different processes seem to be enacted by commercial teams. It had been previously established that individuals involved in such forms of (vertical) communication are often perceived by their peers as knowledge hubs, and powerful team players (Zhu 1996). In fact, such responsibilities and behaviours are often associated with formal software project leadership or individuals occupying coordination and planning related roles (Andre et al. 2011).

Results in this work reveal that team leads and programmers provided context awareness for the other team members and acted as their teams' main information resource (e.g., as evident in the measures for Information sharing, Discussion and Scaffolding). Such behaviours are typically associated with highly skilled roles (Belbin's (2002) IM, PM, SP and RI roles); or with those individuals that are extremely creative, imaginative and insightful (Belbin 2002). Those who communicate more are also generally more aware. This finding for team leads' dominance coincides with those that we identified previously using psycholinguistics (Licorish and MacDonell 2013). While these finding are understandable (particularly for coordination and planning) for those occupying the team lead role given their assigned responsibilities (leading, planning and integration), such skills are not typically required of those who are formally assigned to the programmer role (who were typically expected to contribute to the architecture and code of a component), suggesting that these members were indeed successful at self-organising and adopting informal roles (Hoda et al. 2010). This finding has implications for AGSD, and particularly for instances when programmers, or those occupying similar non-leadership positions, may not be willing to assume such responsibilities (Moe et al. 2008). Additionally, in comparing the outcomes of this role examination to previous literature, it is noted that previous studies have speculated that programmers require fewer communication-related abilities (Acuna et al. 2006a). However, the evidence reported here is divergent to these views. Results in this work indicate that *all* ISD practitioners may actively participate in communication and coordination to enhance their teams' self-organisation if/when the project environment is supportive, or demands such participation. Thus, formal role assignment may not be a sufficient indicator of the need for communication and coordination during AGSD projects.

The pattern of behaviours exhibited by programmers in this work may not be default behaviours, however. While these practitioners may feel a sense of obligation to their teams, a facilitating organisation and work structure may be a prerequisite for encouraging programmers to work across roles as the need arises. Given the evidence uncovered in this work, it is posited that IBM Rational is one such organisation that encourages team members' performance based on their natural abilities, and that promotes non-hierarchical and informal work structures. Such configurations have long been shown to encourage tacit knowledge sharing and cross-fertilization among team members, allowing team members to adapt and execute their tasks based on work demands (Powell 1990). These environments are well suited for ADSD teams, and should be encouraged if such teams are to succeed.

## 6. CONCLUSIONS AND LIMITATIONS

Agile proponents have stressed the need for self-organisation during agile software development to enhance the likelihood of team success. In particular, the ability to self-organise is held to be critical in AGSD settings. Our results show that team leads and programmers were integral to the self-organisation processes of the teams we studied. We contend that the evidence of the way programmers in this work adopted other roles was linked to a facilitating organisation and work structure, and this may be a prerequisite for self-organisation.

We acknowledge that there are limitations to this study, and to the generalisation of our results. In particular, the messages from the three teams may not necessarily represent all IBM Rational Jazz teams' processes in the repository, and the work processes and work culture at IBM are specific to that organisation and may not be representative of organisation dynamics elsewhere. Additionally, our analyses did not consider the background of the practitioners (e.g., period of employment in given roles), which may influence the interaction behaviours these members displayed. Future works are encouraged to consider and address these limitations in designing replication studies.

## REFERENCES

Abreu, R., and Premraj, R. 2009. "How Developer Communication Frequency Relates to Bug Introducing Changes," in: *joint international and annual ERCIM workshops on IWPSE and Evol.* Amsterdam, The Netherlands: ACM, pp. 153-158.

Acuna, S.T., Juristo, N., and Moreno, A.M. 2006a. "Emphasizing Human Capabilities in Software Development," *IEEE Software* (23:2), March-April, 2006, pp 94-101.

Acuna, S.T., Juristo, N., and Moreno, A.M. 2006b. "Emphasizing Human Capabilities in Software Development.," *IEEE Software* (23:2), March-April, 2006, pp 94 - 101.

Andre, M., Baldoquin, M.G., and Acuna, S.T. 2011. "Formal Model for Assigning Human Resources to Teams in Software Projects," *Information and Software Technology* (53:3), pp 259-275.

Arora, A., and Gambardella, A. 2005. "The Globalization of the Software Industry: Perspectives and Opportunities for Developed and Developing Countries," NBE-Research, Washington, DC, pp. 1 - 32.

Bachmann, A., and Bernstein, A. 2009. "Software Process Data Quality and Characteristics: A Historical View on Open and Closed Source Projects," in: *joint international and annual ERCIM workshops on IWPSE and Evol*. Amsterdam, Netherlands: ACM, pp. 119-128.

Belbin, R.M. 2002. *Management Teams: Why They Succeed or Fail*. Woburn, UK: Butterworth-Heinemann.

Bird, C., Nagappan, N., Devanbu, P., Gall, H., and Murphy, B. 2009. "Does Distributed Development Affect Software Quality? An Empirical Case Study of Windows Vista," in: *Proceedings of the 31st ICSE*. Vancouver, BC: IEEE Computer Society, pp. 518-528

Boehm, B.W., and Turner, R. 2003. "Using Risk to Balance Agile and Plan-Driven Methods," *IEEE Journal* (36:6), June 2003, pp 57-66.

Cataldo, M., Bass, M., Herbsleb, J.D., and Bass, L. 2007. "On Coordination Mechanisms in Global Software Development," in: *Second IEEE ICGSE, 2007*. Munich, Germany: IEEE Computer Society, pp. 71-80.

Cohen, J. 1988. *Statistical Power Analysis for the Behavioral Sciences*, (2nd ed.). NJ: Lawrence E Assoc.

Crowston, K., Li, Q., Wei, K., Eseryel, Y.U., and Howison, J. 2007. "Self-Organization of Teams for Free/Libre Open Source Software Development," *Information and Software Technology* (49:6), pp 564-575.

Frost, R. 2007. "Jazz and the Eclipse Way of Collaboration," *IEEE Software* (24:6), pp 114-117.

Glaser, B.G., and Strauss, A.L. 1967. *The Discovery of Grounded Theory: Strategies for Qualitative Research*. Chicago: Aldine Publishing Company.

Gorla, N., and Lam, Y.W. 2004. "Who Should Work with Whom?: Building Effective Software Project Teams," *Commun. ACM* (47:6), pp 79-82.

Henri, F., and Kaye, A.R. 1992. "Computer Conferencing and Content Analysis," in: *Collaborative Learning through Computer Conferencing: The Najaden Papers*. New York: Springer-Verlag, pp. 117-136.

Herbsleb, J.D., and Mockus, A. 2003. "An Empirical Study of Speed and Communication in Globally Distributed Software Development," *IEEE Transactions on Software Engineering* (29:6), pp 481-494.

Highsmith, J. 2004. *Agile Project Management: Creating Innovative Products*. Boston, MA: Pearson Inc.

Hoda, R., Noble, J., and Marshall, S. 2010. "Organizing Self-Organizing Teams," in: *Proceedings of the 32nd ACM/IEEE ICSE - Volume 1*. Cape Town, South Africa: ACM, pp. 285-294.

Holsti, O.R. 1969. *Content Analysis for the Social Sciences and Humanities*. Reading, MA: Addison Wesley.

Hsieh, H.-F., and Shannon, S.E. 2005. "Three Approaches to Qualitative Content Analysis," *Qualitative Health Research* (15:9), November 1, 2005, pp 1277-1288.

Kamaruddin, N.K., Arshad, N.H., and Mohamed, A. 2012. "Chaos Issues on Communication in Agile Global Software Development," in: *2012 IEEE BEIAC*. Kuala Lumpur: IEEE Computer Society, pp. 394-398.

Kampenes, V.B., Dybå, T., Hannay, J.E., and Sjøberg, D.I.K. 2007. "A Systematic Review of Effect Size in Software Engineering Experiments," *Information and Software Technology* (49:11–12), pp 1073-1086.

Layman, L., Williams, L., Damian, D., and Bures, H. 2006. "Essential Communication Practices for Extreme Programming in a Global Software Development Team," *I&ST* (48:9), pp 781-794.

Lee, S., and Yong, H.-S. 2010. "Distributed Agile: Project Management in a Global Environment," *Empirical Software Engineering* (15:2), pp 204-217.

Licorish, S.A., and MacDonell, S.G. 2013. "How Do Globally Distributed Agile Teams Self-Organise? Initial Insights from a Case Study," in: *8th ENASE*. Angers, France: SCITEPRESS, pp. 227-234.

Moe, N.B., Dingsoyr, T., and Dyba, T. 2008. "Understanding Self-Organizing Teams in Agile Software Development," in: *Proceedings of the 19th ASWEC*. Perth, WA: IEEE Computer Society, pp. 76-85.

Oreg, S., and Nov, O. 2008. "Exploring Motivations for Contributing to Open Source Initiatives: The Roles of Contribution Context and Personal Values," *Computers in Human Behavior* (24:5), pp 2055-2073.

Powell, W.W. 1990. "Neither Market nor Hierarchy: Network Forms of Organization. In B. M. Staw and L. L. Cummints, Eds.," *Research in Organizational Behavior* (12), pp 295-336.

Pressman, R.S. 2009. *Software Engineering: A Practitioner's Approach*, (7th ed.). New York: McGraw-Hill.

Serce, F.C., Alpaslan, F.-N., Swigger, K., Brazile, R., Dafoulas, G., Lopez, V., and Schumacker, R. 2009. "Exploring Collaboration Patterns among Global Software Development Teams," in: *Proceedings of the 2009 Fourth IEEE ICGSE*. Limerick, Ireland: IEEE Computer Society.

Sharp, V. 1979. *Statistics for the Social Sciences*. the University of Michigan.

Tjosvold, D. 2008. "The Conflict-Positive Organization: It Depends Upon Us," *JOB* (29:1), pp 19-28.

Zhu, E. 1996. "Meaning Negotiation, Knowledge Construction, and Mentoring in a Distance Learning Course," *Selected Research and Development Presentations at NCAECT*, Indianapolis: ERIC, pp. 821-844.

## ACKNOWLEDGEMENTS

We thank IBM for granting us access to the Jazz repository. Thanks also to the coders for their help. S. Licorish is supported by an AUT VC Doctoral Scholarship Award.